\documentclass[onecolumn, draftclsnofoot, 11pt]{IEEEtran}
\usepackage{amsmath}
\usepackage{amssymb}
\usepackage{amsfonts}
\usepackage{algorithm}
\usepackage{graphicx}
\usepackage{epsfig}
\usepackage{subfigure}
\usepackage{psfrag}
\usepackage{xcolor}
\usepackage{cite}
\usepackage[colorlinks,linkcolor=black,anchorcolor=black,citecolor=black,hyperfootnotes=true]{hyperref}

\title{Sparse Signal Processing for Grant-Free Massive Connectivity: A Future Paradigm for Random Access Protocols in the Internet of Things}
\author{Liang Liu, Erik G. Larsson, Wei Yu, Petar Popovski, \\ \v{C}edomir Stefanovi\'{c}, and Elisabeth De Carvalho}

\begin{document}
\maketitle \thispagestyle{empty} \vspace{-0.7in}

\newtheorem{definition}{Definition}
\newtheorem{assumption}{Assumption}
\newtheorem{lemma}{\underline{Lemma}}
\newtheorem{example}{Example}
\newtheorem{theorem}{Theorem}
\newtheorem{proposition}{Proposition}
\newtheorem{conjecture}{Conjecture}
\newtheorem{remark}{Remark}
\newcommand{\mv}[1]{\mbox{\boldmath{$ #1 $}}}

\begin{abstract}
The next wave of wireless technologies will proliferate in connecting sensors, machines, and robots for myriad new applications, thereby creating the fabric for the Internet of Things (IoT). A generic scenario for IoT connectivity involves a massive number of machine-type connections. But in a typical application, only a small (unknown) subset of devices are active at any given instant, thus one of the key challenges for providing massive IoT connectivity is to detect the active devices first and then to decode their data with low latency. This article advocates the usage of grant-free, rather than grant-based, random access scheme for the problem of massive IoT access and outlines several key signal processing techniques that promote the performance of the considered grant-free strategy, focusing primarily on advanced compressed sensing technique and its application for efficient detection of the active devices. We show that massive multiple-input multiple-output (MIMO) is especially well-suited for massive IoT connectivity in the sense that the device detection error can be driven to zero asymptotically in the limit as the number of antennas at the base station goes to infinity by using the multiple-measurement vector (MMV) compressed sensing techniques. The paper also provides a perspective on several related important techniques for massive access, such as embedding of short messages onto the device activity detection process and the coded random access.
\end{abstract}

\section{Introduction}

\begin{figure}[t]
  \centering
  \includegraphics[width=14cm]{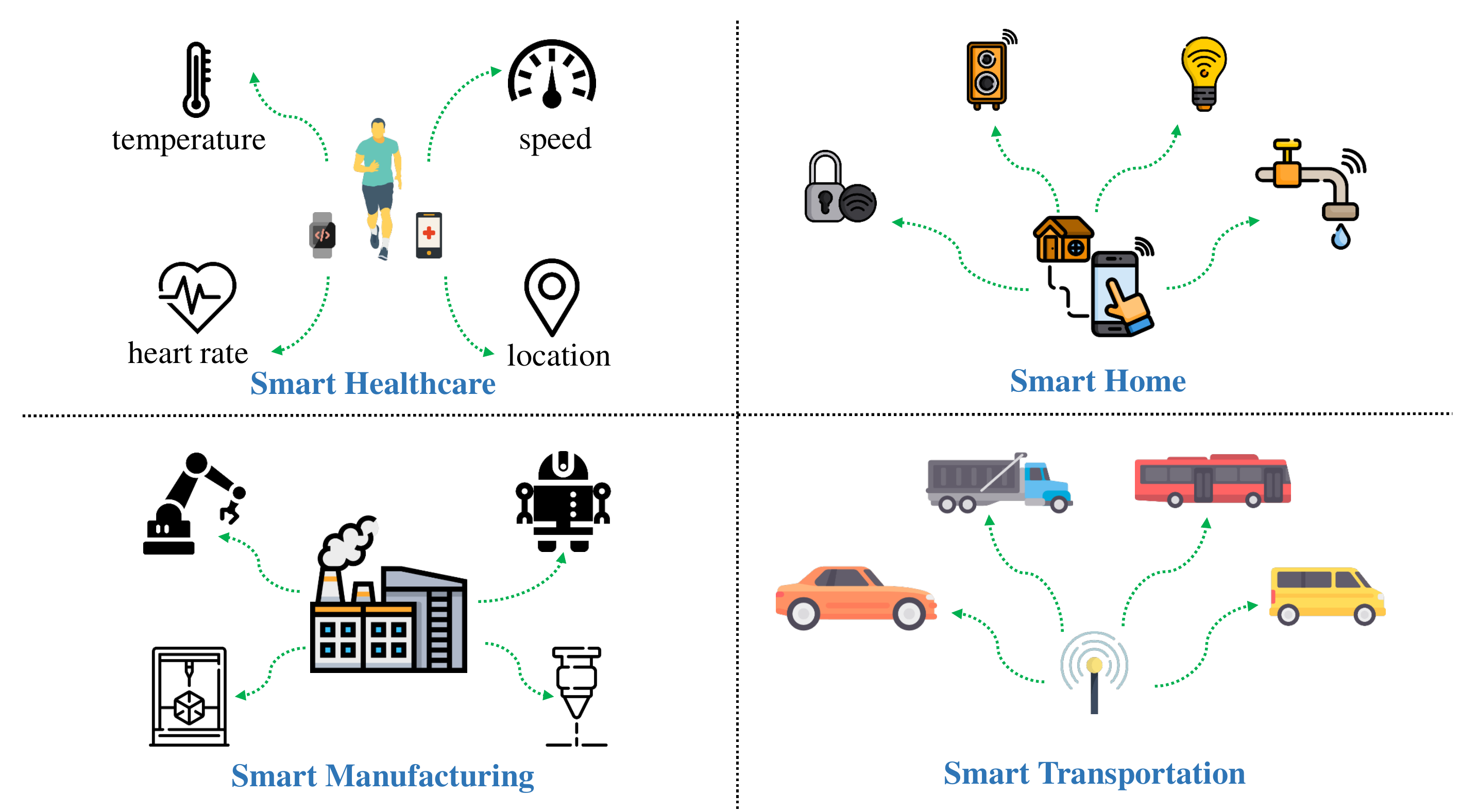}
  \caption{Applications of IoT.}\label{IoT}
\end{figure}

Last decades witnessed the remarkable achievements of the wireless technologies towards offering connectivity to the people. Recently there has been a growing interest to provide ubiquitous connectivity for machines and objects, many of which do not require interactions with humans \cite{Petar16}. This is being driven by the rapid advance of Internet of Things (IoT) that will significantly benefit the way we conduct business, deliver education, health care and government services, and the way we live everyday lives \cite{IoT1}. Typical IoT applications, as shown in Fig. \ref{IoT}, include smart healthcare in which the wearable devices transmit continuous streams of accurate data to the cloud for better care decisions, smart home that enables home automation with the aid of intelligent appliances such as the smart speaker even when the people are at remote locations, smart manufacturing that supports streamlined business operations and optimized productivity in factories via automatically collecting and analyzing data from the sensors for making better-informed decisions to the actuators such as robotics, smart transportation in which the connected vehicles make transportation itself more efficient and help us get from place to place more quickly, and etc. Targeting at the above growing market of IoT, the 5G cellular technologies roadmap has already identified massive machine-type communications (mMTC) as one of the three main use cases, along with enhanced mobile broadband (eMBB) and ultra-reliable and low latency communications (URLLC).

\begin{figure}[t]
  \centering
  \includegraphics[width=12cm]{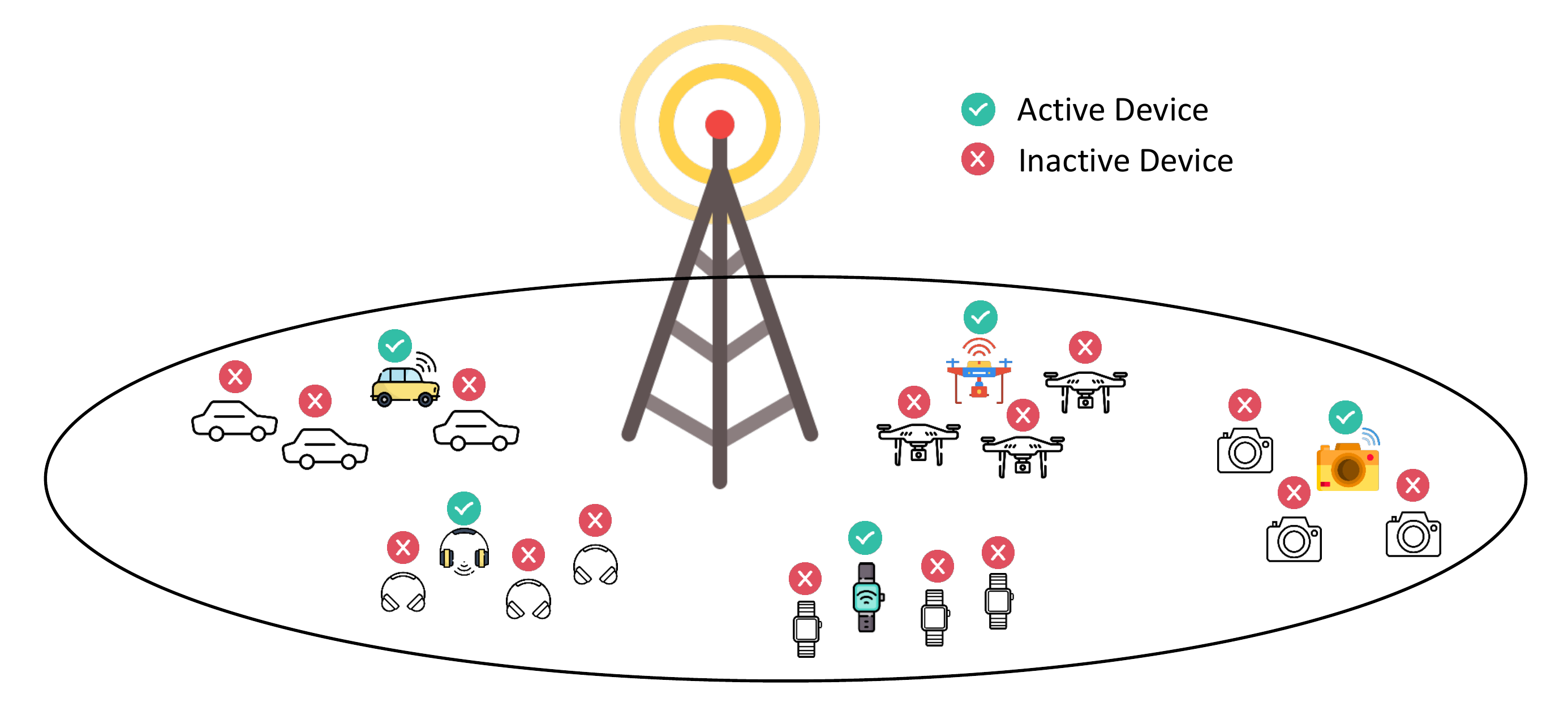}
  \caption{A typical IoT network with a massive number of devices, e.g., drones, smart watches, etc. Arising from the sporadic IoT data traffic, at each time slot only a subset of devices in the network are active.}\label{IoT1}
\end{figure}

The fundamental challenge of mMTC for IoT is to enable
data transmission from a massive number of devices in an efficient and timely manner. However, the key characteristic of the IoT traffic is that the device activity patterns are typically \emph{sporadic}, so that at any given time only a small
and random fraction of all devices are active, as shown in Fig. \ref{IoT1}. The sporadic traffic pattern may be, e.g., due to the fact that often
devices are designed to sleep most of the time in order to save energy
and are activated only when triggered by external events, as is typically
the case in a sensor network. In these scenarios, the active users need to be dynamically identified along with the reception of their data, which is a challenging task.

\subsection{Grant-Based Random Access Scheme}

\begin{figure}[t]
  \centering
  \includegraphics[width=10cm]{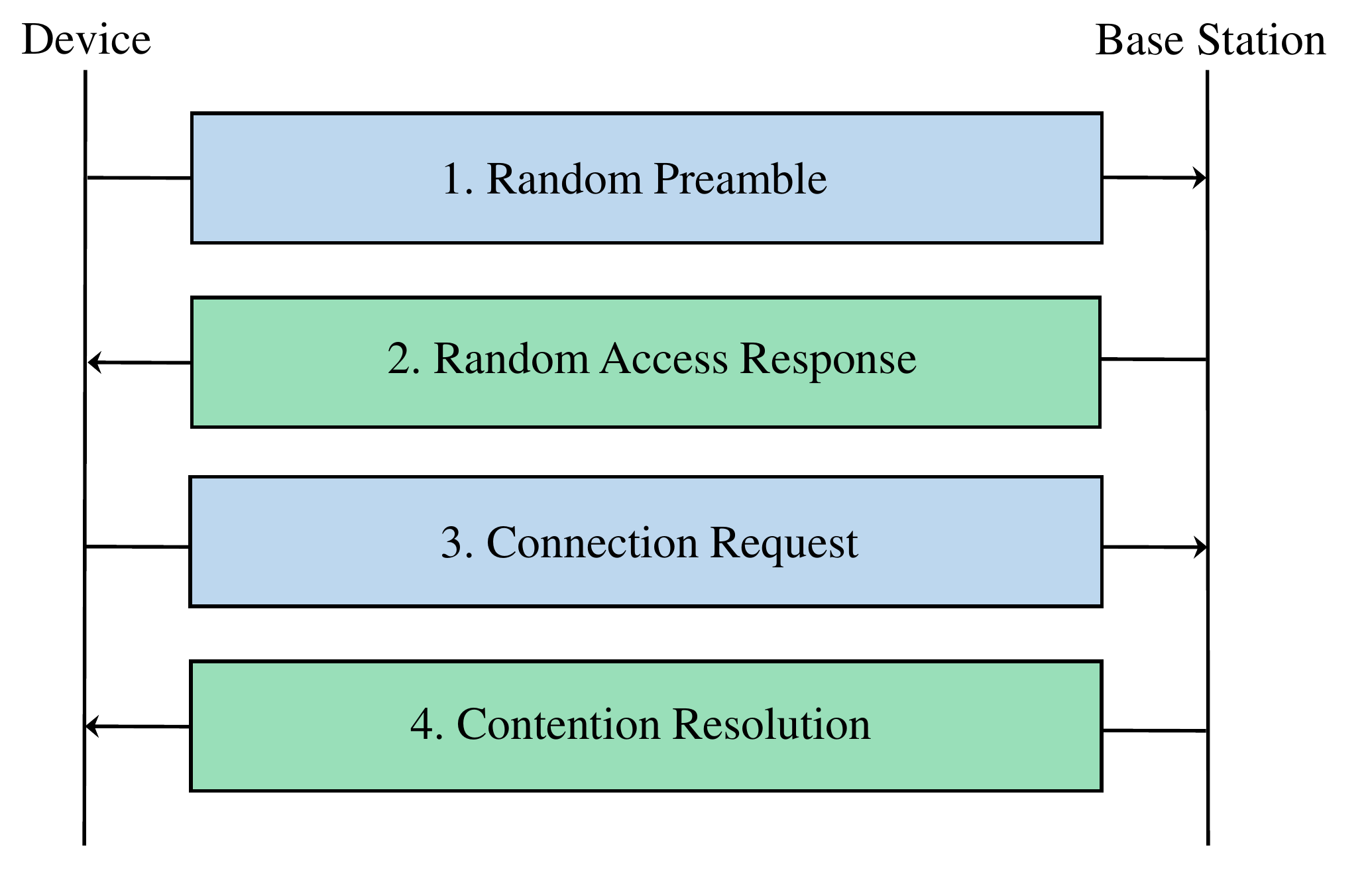}
  \caption{Grant-based random access procedure: due to the lack of coordination, collisions occur when two or more devices select the same pilot and new access attempts may be needed in this case.}\label{Random Access}
\end{figure}

The common user access approach in cellular systems is to perform grant-based random access using the dedicated random-access control channel, so that the uncoordinated devices can contend for physical-layer resource blocks for data transmission \cite{Niyato13}, as shown in Fig.~\ref{Random Access}. Specifically, in the first stage each active device picks a random preamble, sometimes referred to as a pilot sequence, from a predefined set of orthogonal preamble sequences, to notify the base station (BS) that the user has become active. In the second stage, the BS sends a response corresponding to each activated preamble as a grant for transmitting in the next step. In the third stage, each device that has received a response to its preamble transmission sends a connection request in order to demand resources for subsequent data transmission. In case a preamble has been selected by a single device, the connection request of the device is granted by the BS, which in turn sends contention resolution message informing device about the resources reserved for the pending data transmission.
However, if two or more devices have selected the same preamble in the first stage, their connection requests collide. When the BS detects a collision, it does not reply with a contention resolution message; the affected devices restart the random access procedure after a timer expires.
In the above procedure, the messages sent by the active devices in the first and third phases correspond to metadata, since they belong to control information for establishing the connection without containing any data information.

This access mechanism can be seen as an instance of the classical ALOHA, imposing a limit on the number of active devices that can get the grant to access the network.
Recently, extensive efforts have been devoted to different variants of the random access schemes with advanced contention resolution strategies \cite{Caire16,Bjornson}. However, due to the large number of collisions in the massive IoT scenarios, still many users cannot access the network even if some of the colliding connection requests could be resolved, as shown in the following example.

\begin{figure}[t]
  \centering
  \includegraphics[width=10cm]{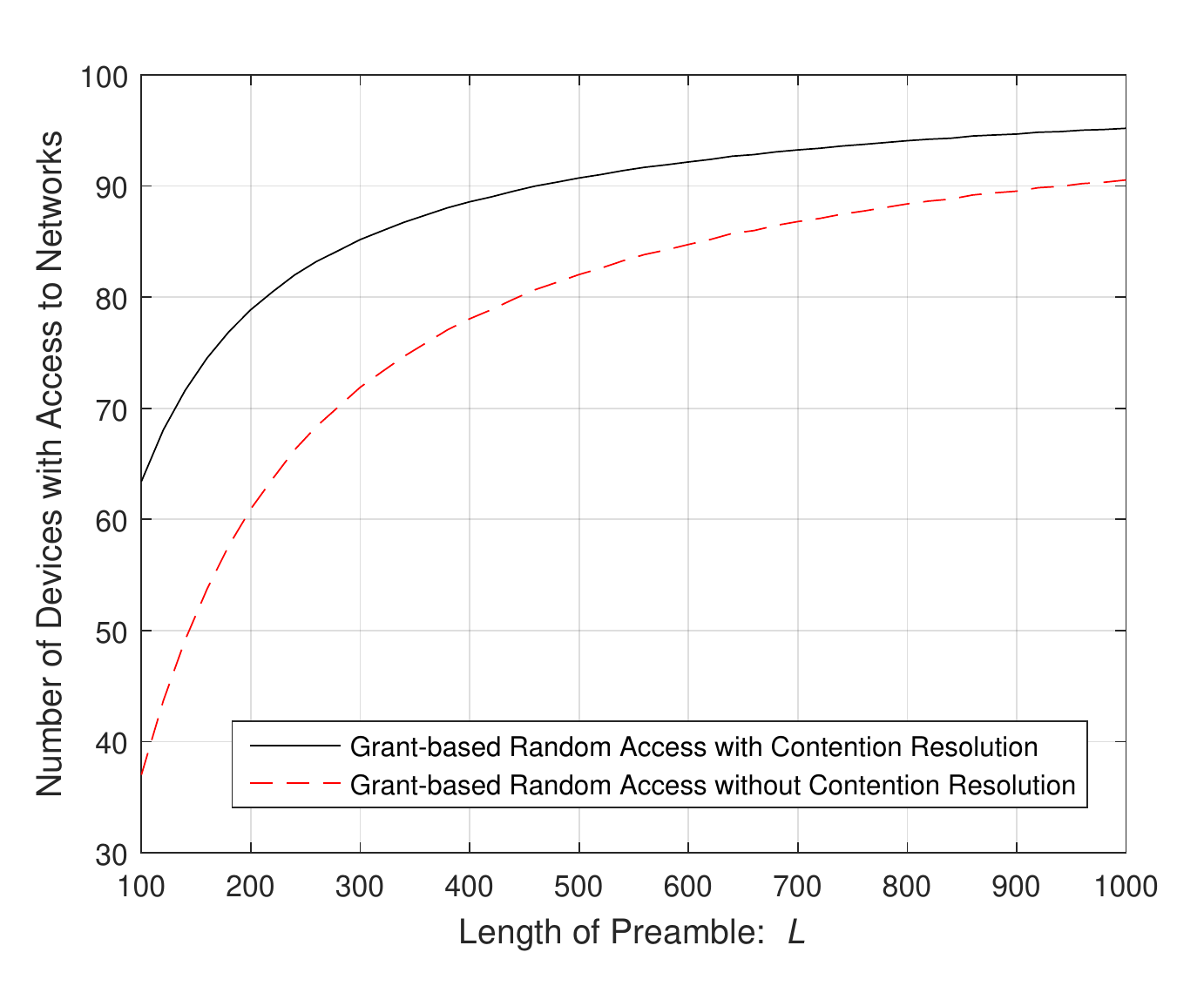}
  \caption{Grant-based random access with orthogonal pilots and capture: How many users can access the network?}\label{Number_of_Successful_Users}
\end{figure}

\begin{example}\label{example1}
Consider a cellular network consisting of one BS and $2000$ users. Let $L$ denote the length (and thus the number) of the orthogonal preambles available for the devices to choose. Assume that in each time slot, $100$ out of these $2000$ devices are active, each picking one of the $L$ orthogonal pilots at random. The coherence bandwidth and the coherence time of the wireless channel are 1MHz and 1ms, respectively, thus in each coherence block 1000 symbols can be transmitted. Moreover, we assume both the scenario in which the contention resolution is not performed and the scenario in which even when there is a collision, the BS can always grant access to one of the colliding devices\footnote{This could happen, e.g., due to the \emph{capture effect}, in random access networks.}. Under this setup, the average numbers of devices that are granted the permission to access the network, for both the cases with and without contention resolution, versus different values of $L$ are plotted in Fig. \ref{Number_of_Successful_Users}. The plot is obtained by Monte Carlo simulations. It is observed that to guarantee $90\%$ success rate, at least $L=470$ and $L=930$ out of $1000$ symbols are needed, respectively, as pilot for the cases with and without contention resolution.
\end{example}

A question arising from the above example is how we can accommodate more devices with low latency requirement in the future massive IoT connectivity systems. One promising solution is the \emph{grant-free} random access scheme based on the advanced \emph{compressed sensing} techniques.

\subsection{Grant-Free Random Access Scheme}
\begin{figure}[t]
  \centering
  \includegraphics[width=10cm]{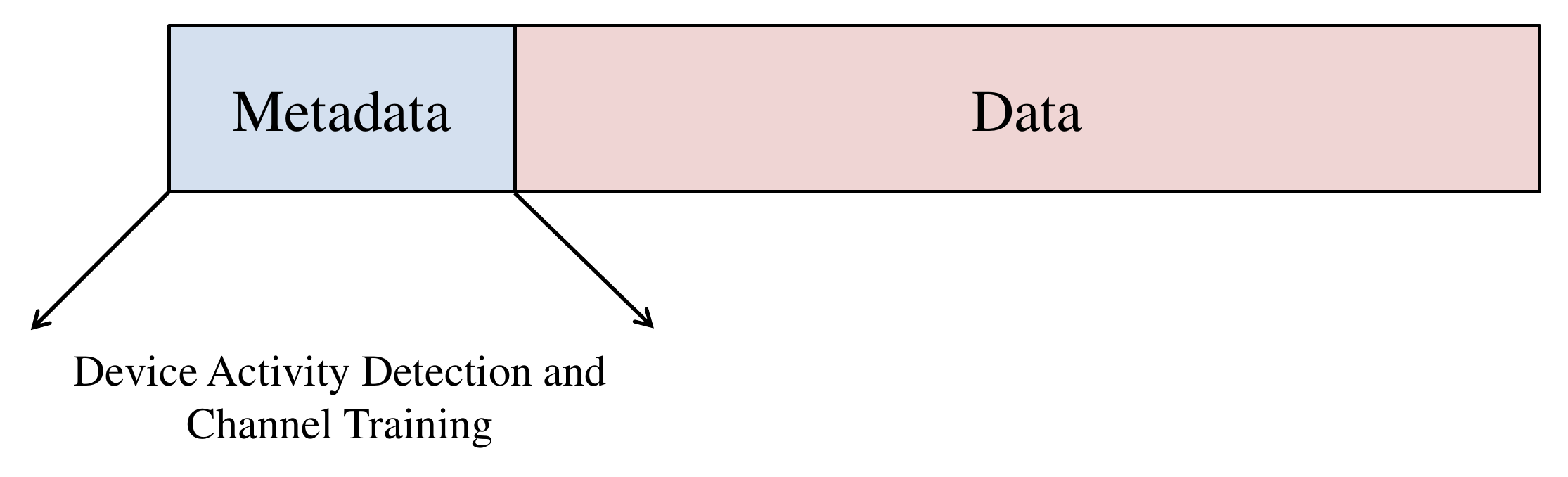}
  \caption{Grant-free transmission strategy: Metadata contains preamble for device activity detection and channel estimation, and data is directly transmitted after metadata without waiting for the grant from BS.}\label{Grant Free}
\end{figure}
Under the grant-free random access scheme, each active device directly transmits its metadata and data to the BS without waiting for any permission, as shown in Fig. \ref{Grant Free}. Specifically, in contrast to the grant-based random access scheme in which pilot sequences are randomly selected at each time slot, each device under the grant-free random access scheme is pre-assigned with a unique pilot sequence used for all the time slots. This pilot sequence thus also serves as the ID for this user and is reminiscent of the role that code-division multiple-access (CDMA) sequence plays in facilitating the extraction of a user data under interference from other users. At each time slot, the BS first detects the active devices by detecting which pilot sequences are used. Next, the BS estimates their channels based on the received metadata, and then decodes the data with the estimated channels \cite{Liu17_massive1,Liu17_massive2}.

Obviously, the very fact that both metadata and data in the grant-free access are sent in a single step offers the possibility to decrease the access latency compared to the grant-based access.
However, device activity detection is now more challenging, because due to the massive number of devices in the network as well as the limited channel coherence time, it is not possible to assign orthogonal pilot sequences to all the devices. The difference with the classical CDMA systems is that the activation dynamics covers a much larger population, placing this problem in the realm of sparse signal processing.

This article aims to pave the way for a theoretical investigation on how the sparse signal processing technologies can enable accurate and efficient active device detection under the grant-free access scheme. We first point out that the device activity detection can be cast into a compressed sensing problem. Next, a random pilot sequence design is introduced, and the use of approximate message passing (AMP) algorithm \cite{donoho_amp} is proposed for detecting the active devices. Further, we show that massive multiple-input multiple-output (MIMO) \cite{Marzetta10,larsson14}, which has already exhibited outstanding performance for enhancing the spectrum efficiency in human-type communications, provides an opportunity to leverage the so-called \emph{multiple-measurement vector} (MMV) compressed sensing technique \cite{schniter_mmv,Ye11} to achieve asymptotically perfect device activity detection accuracy in the massive IoT machine-type communications. Another important fact about mMTC is that it dominantly relies on short packet transmissions. We elaborate on a new method to embed a small number of information bits in the short packets that can be decoded in the device activity detection process. This is enabled by letting each active device randomly select one pilot from a predefined set and letting the BS detect which pilot is used by each active device using AMP. Finally, this article discusses the related technique of coded ALOHA \cite{PSLP2014} for device activity detection.

\section{Device Activity Detection as a Compressed Sensing Problem}
As discussed in the above, it is the sporadic IoT traffic and device activity detection that impose the greatest challenge in the design of the grant-free device access protocol. Interestingly, it is also the sporadic IoT traffic itself that provides a promising opportunity for tackling this challenge. As only a small subset of users are active at each time slot, user activity detection amounts to a sparse signal recovery problem.

Suppose that there are $N$ users in the system, which are denoted by the set $\mathcal{N}=\{1,\cdots,N\}$. Further, assume that the BS is equipped with one antenna, and the channel from user $n$ to the BS is denoted by $h_n$. In each coherent time slot, define the user activity indicator function as
\begin{align}\label{eqn:user activity indicator}
\alpha_n=\left\{\begin{array}{ll}1, & {\rm if ~ user} ~ n ~ {\rm is ~ active}, \\ 0, & {\rm otherwise},\end{array} \right. ~~~ \forall n\in \mathcal{N}.
\end{align}
Assume that each device $n$ decides in each
coherence block whether or not to access the channel with probability
$\epsilon_n$ in an independent manner. Then, $\alpha_n$ can be modeled as a Bernoulli random variable so that
${\rm Pr}(\alpha_n=1)=\epsilon_n$, ${\rm Pr}(\alpha_n=0)=1-\epsilon_n$, $\forall n$. The sparse activity level $\epsilon_n$ depends on the specific applications. The model is sufficiently general so that it can capture a variety of applications, e.g., a sensor fusion network in which the sampling rates at different sensors may even be different.

Suppose that each device $n$ is assigned with one pilot sequence $\mv{a}_n\in \mathbb{C}^{L\times 1}$ with $\|\mv{a}_n\|^2=1$, where $L$ denotes the length of device pilot sequence. Furthermore, we assume that the active users are synchronized within the cyclic prefix, and accurately enough in frequency such that the block fading assumption yields a legitimate model for the channel. This is justified by having the BS send a beacon that invites uplink transmissions from the active devices. The received signal at the BS for device activity detection is then
\begin{align}\label{eqn:received signal single antenna}
\mv{y} =\sqrt{\xi }\sum\limits_{n\in \mathcal{N}}\alpha_nh_n\mv{a}_n+\mv{z}=\sqrt{\xi }\mv{A}\mv{x}+\mv{z},
\end{align}
where $\mv{y}=[y_1,\cdots,y_L]^T \in \mathbb{C}^{L\times 1}$ is the
received signals over $L$ symbols, $\xi$ is the total transmit energy of the pilot for each active device, $\mv{z}\in \mathbb{C}^{L\times 1}\sim \mathcal{CN}(\mv{0},\sigma^2\mv{I})$
is the independent additive white Gaussian noise (AWGN) at the BS, $\mv{A}=[\mv{a}_1, \cdots, \mv{a}_N]$ is the collection of pilot sequences of all the devices, and $\mv{x}=[x_1,\cdots,x_N]^T$ with $x_n=\alpha_n h_n$ denoting the effective channel of device $n$. The goal for the BS is to detect the active devices and detect their channels
by recovering $\mv{x}$ based on the noisy observation $\mv{y}$.

Restricted by the limited coherence time in a practical massive IoT connectivity scenario, the length of device pilot sequence is much smaller than the number of devices, i.e., $L\ll N$. Hence, (\ref{eqn:received signal single antenna}) describes an underdetermined linear system with more unknown variables than equations. However, since $\mv{x}$ is sparse with many zero entries based on (\ref{eqn:user activity indicator}), such a reconstruction problem is a sparse optimization problem that can be possibly solved via non-linear compressed sensing techniques.

There are two main theoretical questions in compressed sensing. First, how to design the sensing matrix $\mv{A}$ so as to capture almost all the information about $\mv{x}$ with a minimal cost $L$? Second, given a sensing matrix $\mv{A}$, how to recover $\mv{x}$ from the noisy observation $\mv{y}$ even if $L<N$? In fact, these two questions are coupled: a good design of the sensing matrix $\mv{A}$ leads to an easier algorithm for recovering the sparse signal $\mv{x}$. For the massive IoT connectivity setting, this indicates that the device pilot sequences should be carefully designed to enable efficient activity detection schemes at the BS side.

Although a number of desirable properties for a good sensing matrix are known, e.g., restricted isometry property (RIP), optimizing the sensing matrix design is quite a challenging problem. This magazine article mainly focuses on simple ways to construct the sensing matrix $\mv{A}$ that are easy to be implemented for practical pilot design. In Sections \ref{sec:AMP based Solution} and \ref{sec:From SMV to MMV Problem: Massive MIMO Technique for Massive IoT Connectivity}, we consider the case when each entry of $\mv{A}$ is i.i.d. randomly generated based on Gaussian distribution and review the AMP algorithm \cite{donoho_amp} to recover $\mv{x}$ \cite{zhilin_TSP,Liu17_massive1}. Later in Section \ref{sec:Other Compressed Sensing Techniques for Device Activity Detection}, we will briefly review the other choices of the sensing matrix $\mv{A}$ and the corresponding compressed sensing algorithms, e.g., the sparse-graph based algorithm with a sparse $\mv{A}$ \cite{Kannan15}, and their applications in device activity detection, e.g., coded slotted ALOHA \cite{PSLP2014}.

\section{AMP-based Device Activity Detection} \label{sec:AMP based Solution}

AMP, proposed in the seminal work \cite{donoho_amp}, is an efficient iterative thresholding method designed for large-scale compressed sensing problems, making it appealing in the massive IoT connectivity scenario of interests. An attractive feature of the AMP framework is that it allows an analytic performance characterization via the so-called \emph{state evolution} \cite{bayati_montanari_amp}. In the following, we introduce how the AMP algorithm works for device activity detection in massive IoT connectivity.

\subsection{Device Pilot Sequence Design}

We assume in this section that the entries of user pilots are generated from i.i.d.\ complex Gaussian distribution with
zero mean and variance $1/L$, i.e.,
\begin{align}\label{eqn:Gaussian pilot}
a_{n,l}\sim \mathcal{CN}(0,1/L), ~~~ \forall n,l.
\end{align}
This particular choice of user pilot sequence is convenient for use with the AMP algorithm for two reasons: first, the convergence of the AMP algorithm for device activity detection is guaranteed if $\mv{A}$ is generated in this way \cite{donoho_amp}; second, with such a Gaussian sensing matrix, the state evolution of the AMP algorithm is well established \cite{bayati_montanari_amp}, based on which detection performance, e.g., \emph{missed detection probability} (probability that an active device is not detected) and \emph{false alarm probability} (probability that an inactive device is declared to be active), can be analytically characterized in the asymptotic limit.

\subsection{Algorithm Design and Performance Analysis}

\subsubsection{General Form of AMP Algorithm}

The AMP algorithm aims to provide an estimate
$\mv{\hat{x}}(\mv{y})$ based on $\mv{y}$ that
minimizes the mean-squared error (MSE)
\begin{align}\label{eqn:MSE}
{\rm MSE} =
\mathbb{E}_{\mv{x}, \mv{y}} || \mv{\hat{x}}(\mv{y}) - \mv{x} ||^2_2.
\end{align}

Based on an approximation of the message passing
algorithm and starting with $\mv{x}^0=\mv{0}$ and $\mv{r}^0=\mv{y}$,
the AMP algorithm proceeds at each iteration
as \cite{donoho_amp,Rangan11}:
\begin{align}
x_n^{t+1} & =\eta_{t,n}((\mv{r}^t)^H\mv{a}_n+x_n^t), \label{eqn:AMP 1 SMV} \\
\mv{r}^{t+1} & =\mv{y}-\mv{A}\mv{x}^{t+1}+\frac{N}{L}\mv{r}^t\sum\limits_{n=1}^N\frac{\eta_{t,n}'((\mv{r}^t)^H\mv{a}_n+x_n^t)}{N}, \label{eqn:AMP 2 SMV}
\end{align}where $t=0,1,\cdots$ is the index of the iteration, $\mv{x}^t=[x_1^t,\cdots,x_N^t]^T$ is
the estimate of $\mv{x}$ at iteration $t$, $\mv{r}^t=[r_1^t,\cdots,r_L^t]^T\in \mathbb{C}^{L\times 1}$
denotes the corresponding residual, $\eta_{t,n}(\cdot):
\mathbb{C}\rightarrow \mathbb{C}$ is the so-called denoiser, and $\eta_{t,n}'(\cdot)$ is the first-order derivative of $\eta_{t,n}(\cdot)$.
The basic intuition is that since the solution should minimize $\|\mv{y}-\mv{A}\mv{x}\|^2$, algorithm makes progress in (\ref{eqn:AMP 1 SMV}) by moving in the direction of the gradient of $\|\mv{y}-\mv{A}\mv{x}^t\|^2$, i.e., $(\mv{r}^t)^H\mv{a}_n$, $n=1,\cdots,N$, and then promotes sparsity by applying an
appropriately designed denoiser $\eta_{t,n}(\cdot)$. The residual is then updated in (\ref{eqn:AMP 2 SMV}), but corrected
with a so-called Onsager term involving $\eta_{t,n}'(\cdot)$.

\subsubsection{State Evolution}

An important analytical result about the AMP algorithm is the so-called state evolution in the asymptotic regime when $L,K,N\rightarrow \infty$, while their ratios converge to some fixed
positive values $N/L \rightarrow \omega$ and $K/N \rightarrow
\epsilon=\lim_{N\rightarrow \infty} \sum_n \epsilon_n/N$ with $\omega, \epsilon \in (0,\infty)$. In systems for massive IoT connectivity, these assumptions indicate that the length of the pilot sequence is in the same order of the number of active users or total users. After the $t$th iteration of the AMP algorithm, define a set of random variables $\hat{X}_n^t$'s as
\begin{align}\label{eqn:Gaussian noise}
\hat{X}_n^t=X_n+\tau_tV_n, ~~~ \forall n,
\end{align}where the random variables $X_n$'s capture the distributions of $x_n$'s, $V_n$ follows the normal distribution, i.e., $V_n\in \mathcal{CN}(0,1)$, and is independent of $X_n$ as well as $V_j$, $\forall j\neq n$, and $\tau_t$ is the state variable, which changes from iteration to iteration as modeled by a simple scalar iterative function known as the MSE map:
\begin{align}\label{eqn:state evolution}
\tau_{t+1}=\frac{\sigma^2}{\xi}+\omega \mathbb{E}[|\eta_{t,n}(X_n+\tau_tV_n)-X_n|^2].
\end{align}Here, the expectation is over the random variables $X_n$'s and $V_n$'s and over all $n$. Under the aforementioned asymptotic regime, \cite{bayati_montanari_amp} shows that applying the denoiser to $(\mv{r}^t)^H\mv{a}_n+x_n^t$ in (\ref{eqn:AMP 1 SMV}) is statistically equivalent to applying the denoiser to $\hat{X}_n^t$ as shown in (\ref{eqn:Gaussian noise}).

The statistical model of AMP as given in (\ref{eqn:Gaussian noise}) and (\ref{eqn:state evolution}) can be utilized to design the denoiser functions $\eta_{t,n}(\cdot)$'s in (\ref{eqn:AMP 1 SMV}) and to quantify the performance of the AMP algorithm.

\subsubsection{Minimax Framework for Denoiser Design}
The flexibility in the AMP algorithm design lies in the denoiser $\eta_{t,n}(\cdot)$ in (\ref{eqn:AMP 1 SMV}). In the AMP literature, the prior distribution of $\mv{x}$ is in general assumed to be unknown. In this case, the denoiser $\eta_{t,n}(\cdot)$ is designed under the minimax framework to optimize the AMP algorithm performance for the worst-case or least-favorable distribution of $\mv{x}$ \cite{Donoho10}. Such a design leads to a soft thresholding denoiser for promoting sparsity even for $\mv{x}$ with the worst-case distribution \cite{donoho_amp}:
\begin{align}\label{eqn:soft thresholding denoise}
\eta_{t,n}(\hat{x}_n^t)=\left(\hat{x}_n^t-\frac{\theta_n^t
\hat{x}_n^t}{|\hat{x}_n^t|}\right)\mathbb{I}(|\hat{x}_n^t|>\theta_n^t),
\end{align}where the distribution of $\hat{x}_n^t$ is captured by $\hat{X}_n^t$, and $\theta_n^t>0$ is the threshold for device $n$ for the $t$th iteration of the AMP algorithm, which can be optimized based on the state evolution (\ref{eqn:state evolution}) to minimize the MSE as given in (\ref{eqn:MSE}). With this denoiser, after the $t$th iteration of the AMP algorithm as shown in (\ref{eqn:AMP 1 SMV}) and (\ref{eqn:AMP 2 SMV}), device $n$ is declared to be active if $|(\mv{r}^t)^H\mv{a}_n+x_n^t|>\theta_n^t$, and declared to be inactive otherwise. Note that AMP with soft thresholding implicitly solves the LASSO problem \cite{Donoho10}, i.e., the sparse signal recovery problem as an $\ell_1$-penalized least squares optimization.

\subsubsection{Bayesian Framework for Denoiser Design}
On the other hand, if the distribution of $\mv{x}$ is known in (\ref{eqn:received signal single antenna}), we can design the minimum mean-squared
error (MMSE) denoiser via the Bayesian approach to minimize the MSE for the estimation of $\mv{x}$ as given in (\ref{eqn:MSE}) \cite{Donoho10}. Considering the equivalent signal model (\ref{eqn:Gaussian noise}), the MMSE denoiser is given as the following conditional expectation
\begin{align}\label{eqn:MMSE denoiser}
\eta_{t,n}(\hat{x}_n^t)=\mathbb{E}[X_n|\hat{X}_n^t=\hat{x}_n^t], ~~~ \forall t, n,
\end{align}where the expectation is over $X_n$ and $\hat{X}_n^t$.

For example, if we assume a Rayleigh fading channel such that $h_n\sim \mathcal{CN}(0,\beta_n)$, where $\beta_n$ denotes the path-loss and shadowing
component of user $n$ and is assumed to be known by the BS, then the effective channel $x_n=\alpha_nh_n$ follows a Bernoulli-Gaussian distribution. Under this particular distribution of $\mv{x}$, an analytical expression of the above MMSE denoiser can be found in \cite{zhilin_TSP}, which is in general non-linear and has complicated form. Similar to the soft thresholding denoiser case, with the MMSE denoiser (\ref{eqn:MMSE denoiser}), we can detect the user activity based on whether the magnitude of $(\mv{r}^t)^H\mv{a}_n+x_n^t$ is larger than or smaller than a carefully designed threshold $\theta_n^t$.

\begin{figure}[t]
   \centering
   \includegraphics[width=10cm]{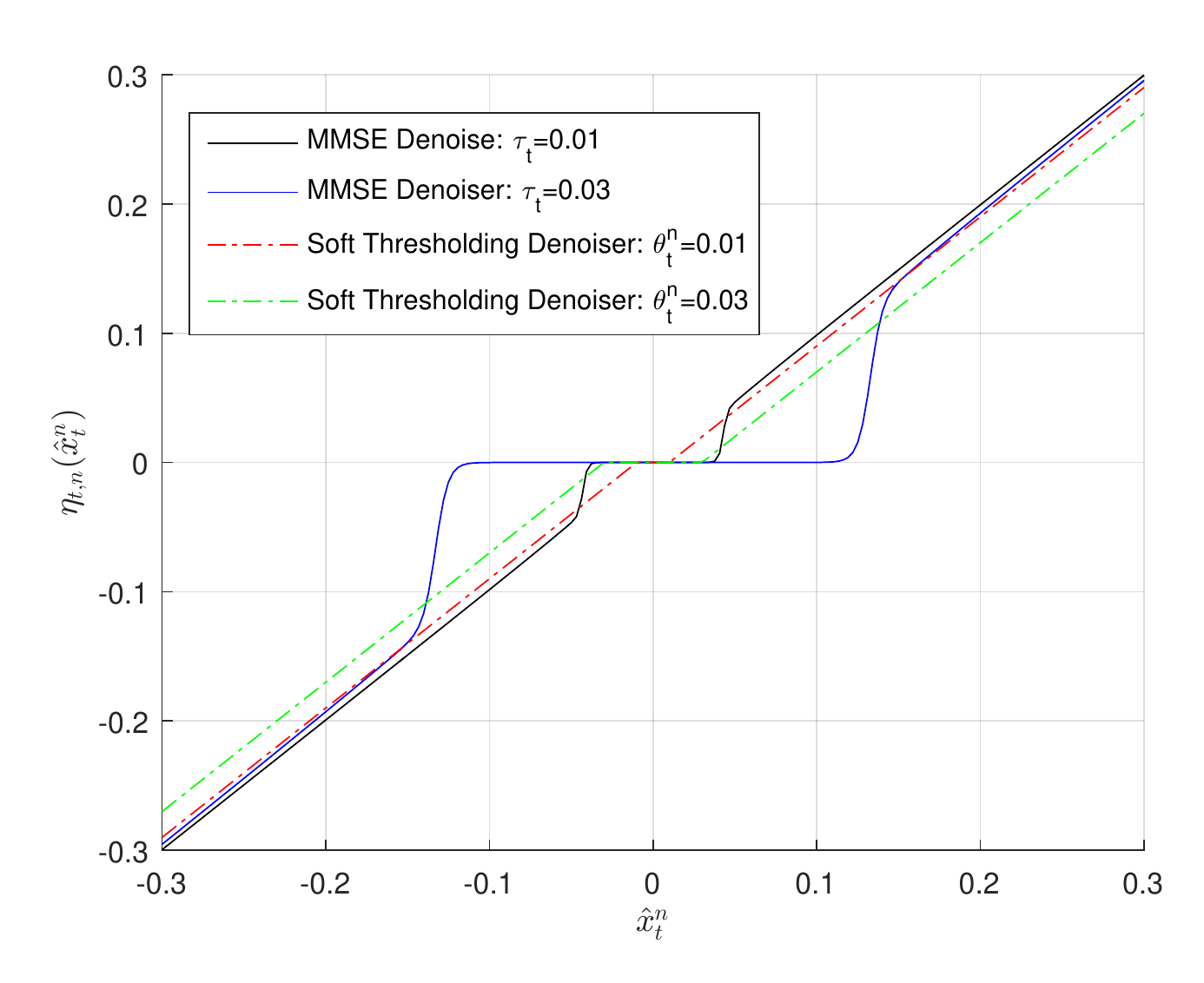}
   \caption{MMSE denoiser versus soft thresholding denoiser in the AMP algorithm.}\label{Denoiser}
 \end{figure}

A comparison between the soft thresholding and MMSE denoisers with a Bernoulli-Gaussian distributed $\mv{x}$ is given in Fig. \ref{Denoiser}. It can be observed that the MMSE denoiser is also a thresholding based denoiser, but more ``soft'' around the regime around the threshold. Moreover, the threshold for the MMSE denoiser is obtained by calculating (\ref{eqn:MMSE denoiser}) to minimize the MSE (\ref{eqn:MSE}), while the design of the threshold for the soft thresholding denoiser follows a minimax framework, which is not optimal given a particular distribution of $\mv{x}$ in general.

\subsubsection{Analytical Performance Characterization}\label{sec:Analytical Performance Characterization}

The state evolution also allows an analytical performance characterization of the AMP algorithm. For example, with both the soft thresholding and MMSE denoisers, missed detection event happens if one user is active but $|(\mv{r}^t)^H\mv{a}_n+x_n^t|<\theta_n^t$, while false alarm event happens if one user is inactive but $|(\mv{r}^t)^H\mv{a}_n+x_n^t|>\theta_n^t$. Since $\hat{x}_n^t$ defined in (\ref{eqn:Gaussian noise}) captures the statistical distribution of $(\mv{r}^t)^H\mv{a}_n+x_n^t$, the probabilities of missed detection and false alarm for device $n$ after the $t$th iteration of the AMP algorithm thus can be expressed as
\begin{align}
& P_{t,n}^{{\rm MD}}={\rm Pr}(\hat{x}_n^t<\theta_n^t|\alpha_n=1), \label{eqn:missed detection} \\
& P_{t,n}^{{\rm FA}}={\rm Pr}(\hat{x}_n^t>\theta_n^t|\alpha_n=0), \label{eqn:false alarm}
\end{align}respectively.

Given the distribution of $\mv{x}$ and denoiser $\eta_{t,n}(\cdot)$, we can track the values of $\tau_t$'s over iterations based on the state evolution (\ref{eqn:state evolution}), then calculate the probabilities of missed detection and false alarm based on (\ref{eqn:missed detection}) and (\ref{eqn:false alarm}).

\begin{figure}[t]
  \centering
  \includegraphics[width=10cm]{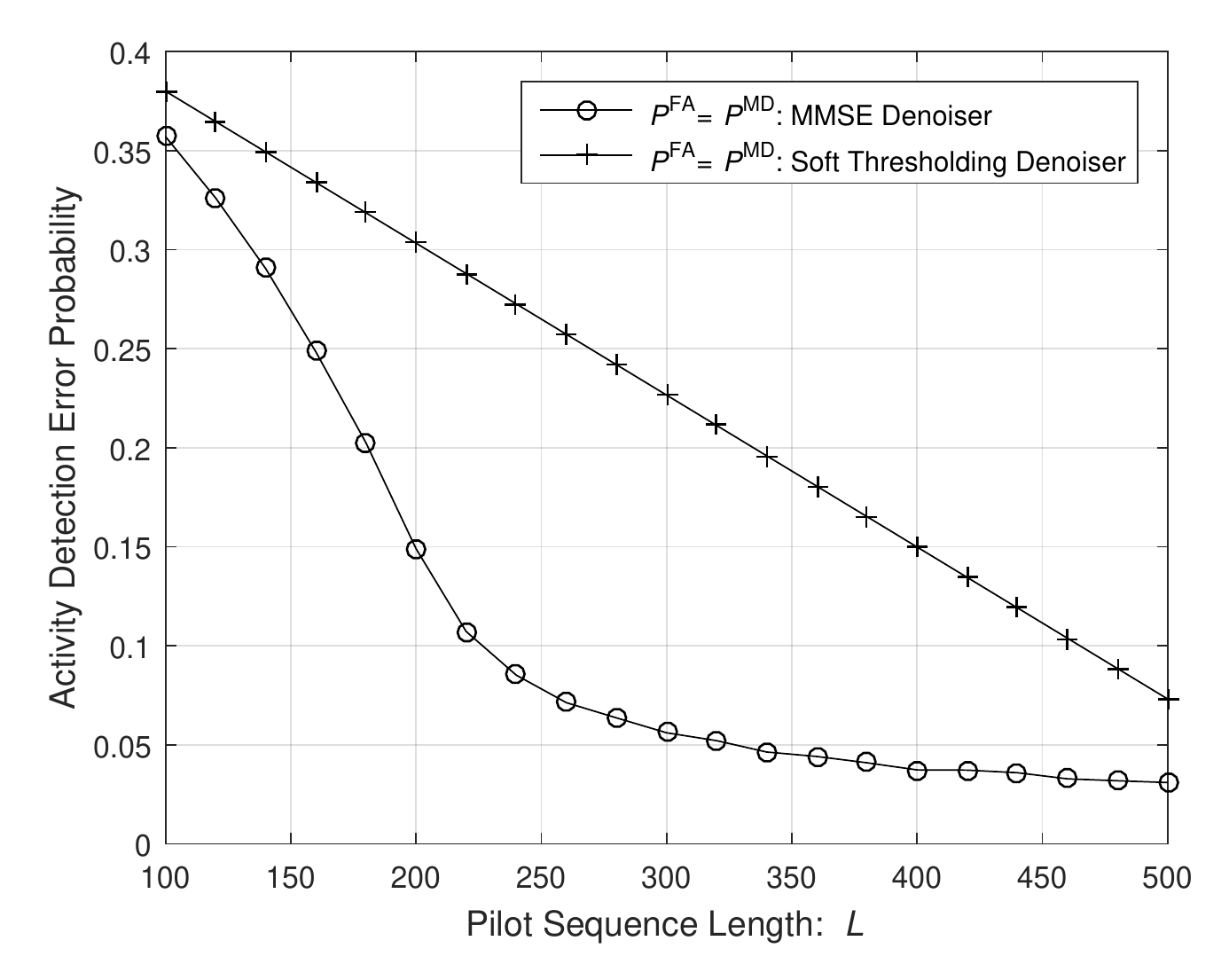}
  \caption{Probabilities of missed detection and false alarm versus pilot sequence length $L$.}\label{MD_FL_SMV}
\end{figure}

\begin{example}\label{example2}
Here we provide a numerical example to show the probabilities of missed detection and false alarm achieved by the AMP algorithm, under the same setup that is used in Example \ref{example1}. The $N=2000$ devices are assumed to be randomly located in a cell with a radius $1000$ meters, while each device accesses the channel with an identical probability $\epsilon_n=0.05$, $\forall n$, i.e., $\epsilon=0.05$ and $K=100$ of the $N=2000$ devices are active at any given time. The transmit
power of each user for sending its pilot
is $\rho^{\rm pilot}=23$dBm. The power spectral density of the AWGN at
the BS is assumed to be $-169$dBm/Hz. Moreover, we define the system-level missed detection and false alarm probabilities as $P^{{\rm MD}}=\sum_{n=1}^NP_n^{{\rm MD}}$ and $P^{{\rm FA}}=\sum_{n=1}^NP_n^{{\rm FA}}$,  where $P_n^{{\rm MD}}$ and $P_n^{{\rm FA}}$ denote the missed detection and false alarm probabilities of device $n$ achieved by AMP after its convergence. Hence, $\epsilon NP^{{\rm MD}}$ and $(1-\epsilon)NP^{{\rm FA}}$ are the average numbers of missed detection and false alarm events at each time slot in a system with $N$ devices. In addition, under both the soft thresholding and MMSE based AMP algorithms, the thresholds $\theta_n^t$'s are carefully selected such that $P^{{\rm MD}}=P^{{\rm FA}}$.

Fig.~\ref{MD_FL_SMV} shows the device activity detection accuracy achieved by the AMP algorithm with the soft thresholding denoiser and the MMSE denoiser. It is observed that with the MMSE denoiser, $90\%$ active devices can be detected with the AMP algorithm when the length of pilot sequence satisfies $L>220$. Recall that in Example 1 of the random access scheme, such a performance can be achieved only when the pilot sequence length is longer than $470$ even if contention resolution is performed. Moreover, with a careful design of the MMSE denoiser, the MMSE denoiser based AMP algorithm outperforms the soft thresholding denoiser based AMP algorithm in terms of device activity detection.
\end{example}

\section{From SMV to MMV: Massive MIMO for Massive IoT Connectivity}\label{sec:From SMV to MMV Problem: Massive MIMO Technique for Massive IoT Connectivity}

As compared to most other applications of compressed sensing such as imaging, a unique and essential opportunity provided by the wireless massive IoT connectivity system design lies in the potential for utilizing the MMV technique for compressed sensing \cite{schniter_mmv}, thanks to the multi-antenna technologies nowadays used ubiquitously in cellular networks. The previous section deals with the application of compressed sensing technique for user activity detection when the BS is equipped with one antenna. In the literature of compressed sensing, the case with one measurement vector is referred to as a single-measurement vector (SMV) problem. Recently, massive MIMO has emerged as a revolutionary technology for dealing with the future data deluge for human-type communications. This section shows that massive MIMO is also a natural solution for accommodating a huge number of IoT devices for the future machine-type communications. From the compressed sensing perspective, device activity detection in massive MIMO systems corresponds to the MMV problem, which generalizes the sparse signal recovery problem to the case with a group of measurement vectors for a group of signal vectors that are assumed to be jointly sparse and share a common support. It is of both theoretical and practical importance to investigate the role of massive MIMO on massive IoT connectivity, which is the aim of this section.

Suppose that the BS is equipped with $M$ antennas. In this case, the channel from user $n$ to the BS is $\mv{h}_n\in \mathbb{C}^{M\times 1}$. Then, the signal model given in (\ref{eqn:received signal single antenna}) is generalized to
\begin{align}\label{eqn:received signal multiple antenna}
\mv{Y} =\sqrt{\xi }\mv{A}\mv{X}+\mv{Z},
\end{align}where $\mv{Y} \in \mathbb{C}^{L\times M}$ is the matrix of
received signals across $M$ antennas over $L$ symbols, $\mv{X}=[\mv{x}_1, \cdots, \mv{x}_N]^T$ with $\mv{x}_n=\alpha_n\mv{h}_n$ denoting the effective channel of user $n$, and $\mv{Z}=[\mv{z}_1,\cdots,\mv{z}_M]$ with
$\mv{z}_m \sim \mathcal{CN}(\mv{0},\sigma^2\mv{I})$, $\forall m$,
is the independent AWGN at the BS.

\begin{figure}[t]
\begin{center}
\subfigure[SMV compressed sensing problem]{\scalebox{0.35}{\includegraphics*{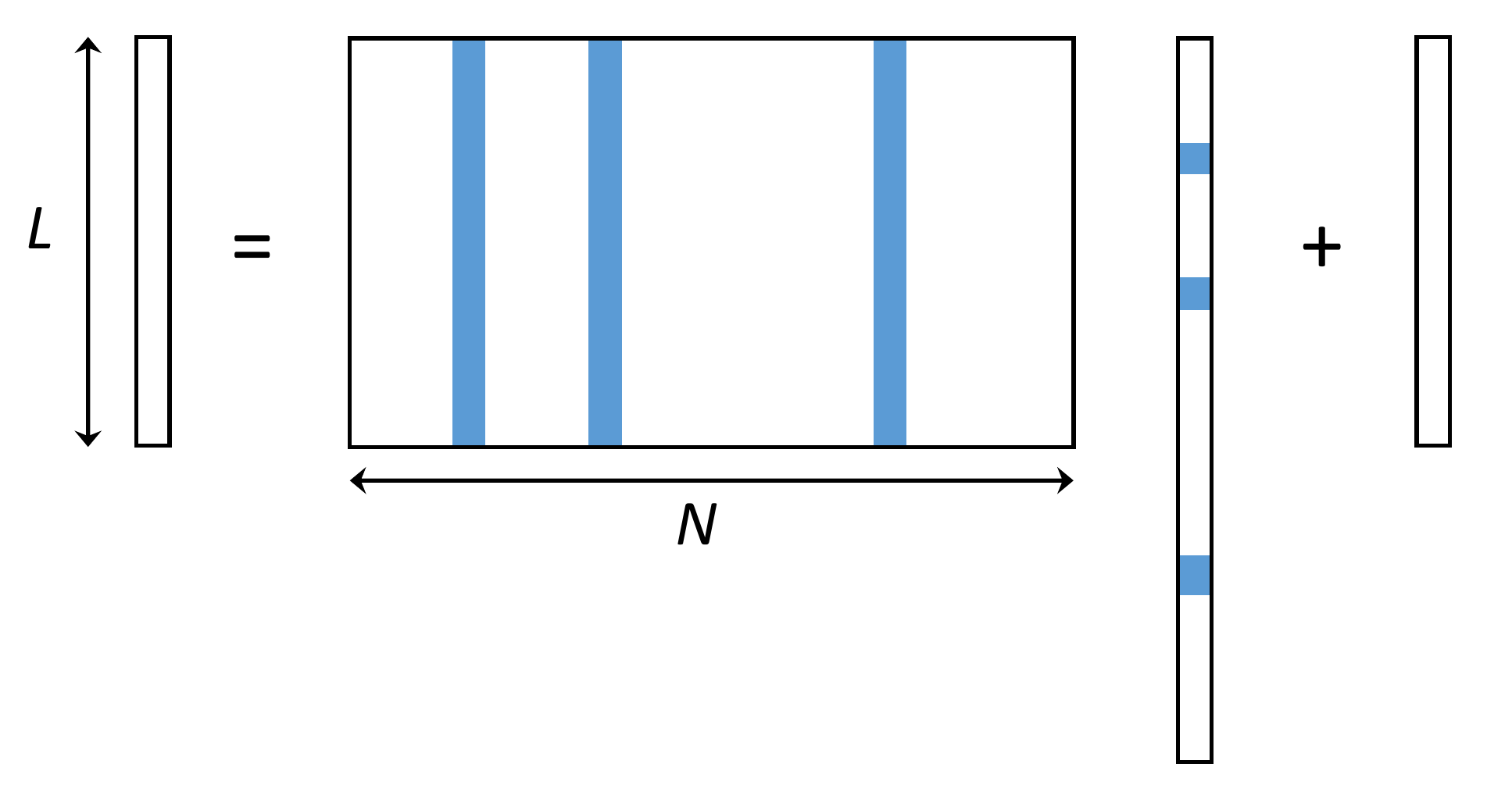}}} \\
\subfigure[MMV compressed sensing problem]{\scalebox{0.35}{\includegraphics*{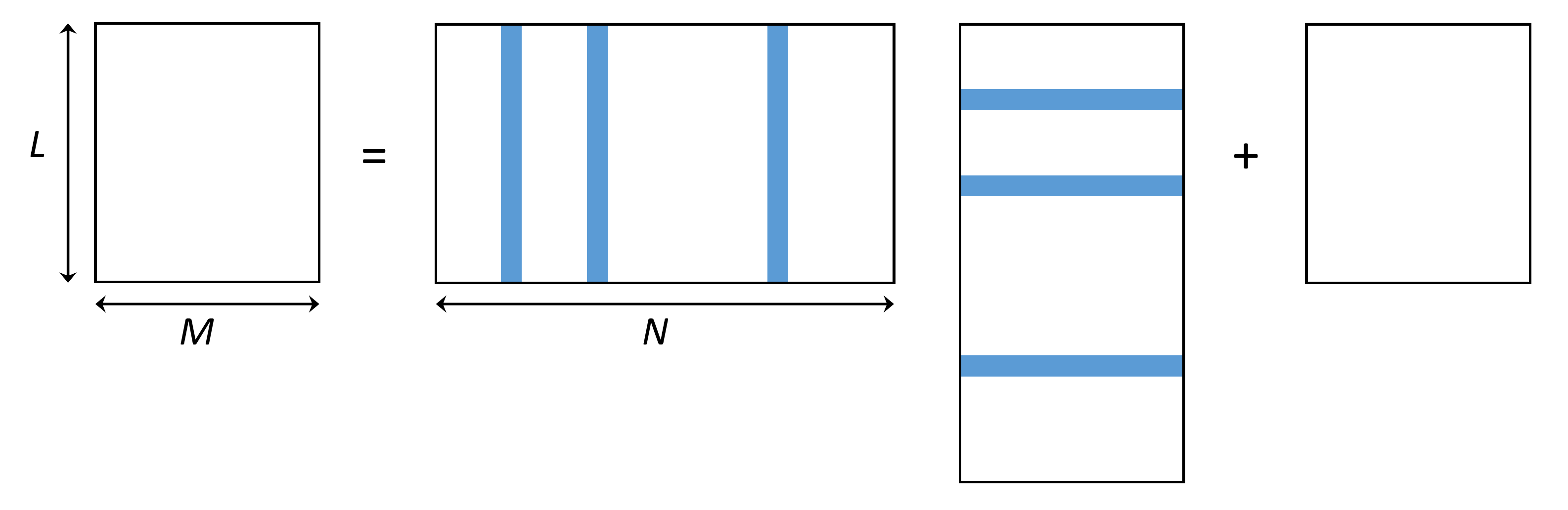}}}
\caption{SMV versus MMV compressed sensing problem: ``group sparsity'' provides additional information about $\mv{X}$ since if one entry is zero, the other entries on the same row should be also zero.}\label{SMV and MMV}
\end{center}
\end{figure}

As compared to the SMV signal model (\ref{eqn:received signal single antenna}), the main difference lies in the fact that $\mv{X}$ in (\ref{eqn:received signal multiple antenna}) is a row-sparse matrix, i.e., if one entry of one particular row of $\mv{A}$ is zero, the other entries of that row must be also zero. This information can be utilized to improve the user detection accuracy. A comparison between the SMV model (\ref{eqn:received signal single antenna}) and MMV model (\ref{eqn:received signal multiple antenna}) is illustrated in Fig. \ref{SMV and MMV}. In the following, we discuss how to generalize the AMP based algorithm in Section \ref{sec:AMP based Solution} to the massive MIMO scenario and to quantify its significant improvement in device activity detection accuracy over the single-antenna BS case.

\subsection{Algorithm Design}

With massive MIMO at the BS, the user pilot sequence assignment still follows (\ref{eqn:Gaussian pilot}), which is the same as the case with one antenna at the BS. However, the AMP algorithm is modified as \cite{Ye11}
\begin{align}
\mv{x}_n^{t+1} & =\eta_{t,n}((\mv{R}^t)^H\mv{a}_n+\mv{x}_n^t), \label{eqn:AMP 1} \\
\mv{R}^{t+1} & =\mv{Y}-\mv{A}\mv{X}^{t+1}+\frac{N}{L}\mv{R}^t\sum\limits_{n=1}^N\frac{\eta_{t,n}'((\mv{R}^t)^H\mv{a}_n+\mv{x}_n^t)}{N}. \label{eqn:AMP 2}
\end{align}As compared to (\ref{eqn:AMP 1 SMV}) and (\ref{eqn:AMP 2 SMV}), the dimensions of the signals are now $\mv{x}_n^t\in \mathbb{C}^{M\times 1}$ and $\mv{R}^t\in \mathbb{C}^{L\times M}$; moreover, the denoiser is a mapping in higher dimension, i.e., $\eta_{t,n}(\cdot):
\mathbb{C}^{M\times 1}\rightarrow \mathbb{C}^{M\times 1}$.

The state evolution of the AMP algorithm still holds for MMV in the asymptotic regime that $L,K,N\rightarrow \infty$ with fixed ratios $N/L\rightarrow \omega$ and $K/N\rightarrow \epsilon$. Specifically, define \cite{Ye11}
\begin{align}\label{eq:stateqmm}
\hat{\mv{X}}_n^t=\mv{X}_n+\mv{\Sigma}_t^{\frac{1}{2}}\mv{V}_n,
\end{align}where the random vector $\mv{X}_n$ captures the distribution of $\mv{x}_n$, $\mv{V}_n\in \mathcal{CN}(\mv{0},\mv{I})$ is the independent Gaussian noise, and $\mv{\Sigma}_t$ can be tracked over iterations as follows
\begin{align}\label{eqn:state evolution MMV}
\mv{\Sigma}_{t+1}=\frac{\sigma^2}{\xi }\mv{I}+
	\omega \mathbb{E}\bigg[(\eta_{t,n}(\mv{X}_n+\mv{\Sigma}_t^{\frac{1}{2}}\mv{V}_n)-\mv{X}_n)(\eta_{t,n}(\mv{X}_n+\mv{\Sigma}_t^{\frac{1}{2}}\mv{V}_n)-\mv{X}_n)^H\bigg].
\end{align}Here, the expectation is over $\mv{X}_n$'s and $\mv{V}_n$'s and over all $n$. Then, in (\ref{eqn:AMP 1}), applying denoiser to $(\mv{R}^t)^H\mv{a}_n+\mv{x}_n^t$ is statistically equivalent to applying denoiser to
\begin{align}\label{eq:stateqmm1}
\hat{\mv{x}}_n^t=\mv{x}_n+\mv{\Sigma}_t^{\frac{1}{2}}\mv{v}_n,
\end{align}where the distributions of $\hat{\mv{x}}_n^t$ and $\mv{v}_n$ are captured by $\hat{\mv{X}}_n^t$ and $\mv{V}_n$.

Based on the above state evolution, denoisers of the MMV-based AMP algorithm can be designed based on different criteria as for the SMV case. For example, the soft thresholding denoiser is
\begin{align}\label{eqn:soft thresholding denoiser MMV}
\eta_{t,n}(\hat{\mv{x}}_n^t)=\left(\hat{\mv{x}}_n^t-\frac{\theta_n^t \hat{\mv{x}}_n^t}{\|\hat{\mv{x}}_n^t\|}\right)\mathbb{I}(\|\hat{\mv{x}}_n^t\|_2>\theta_n^t),
\end{align}Further, assuming Bernoulli Gaussian distributed $\mv{x}_n$'s, the MMSE denoiser \begin{align}\label{eqn:MMSE denoiser MMV}
\eta_{t,n}(\hat{\mv{x}}_n^t)=\mathbb{E}[\mv{X}_n|\hat{\mv{X}}_n^t=\hat{\mv{x}}_n^t],
\end{align}is characterized in \cite{Liu17_massive1}. With both the soft thresholding and MMSE denoisers, after the $t$th iteration of the AMP algorithm, user $n$ can be declared to be active if $\|(\mv{R}^t)^H\mv{a}_n+\mv{x}_n^t\|_2>\theta_n^t$, and declared to be inactive otherwise, where $\theta_n^t$ is the carefully designed threshold for device detection.

\subsection{Asymptotically Perfect Device Activity Detection}

Fix the number of antennas at the BS, $M$, the missed detection and false alarm probabilities from the MMSE denoiser based AMP algorithm, denoted by $P_{t,n}^{{\rm MD}}(M)$ and $P_{t,n}^{{\rm FA}}(M)$ (reducing to (\ref{eqn:missed detection}) and (\ref{eqn:false alarm}) when $M=1$), are characterized in \cite{Liu17_massive1}. Interestingly, perfect device activity detection is achieved in the asymptotic regime of $M\rightarrow \infty$ if the thresholds for device detection, i.e., $\theta_n^t$'s, are properly selected (c.f. \cite[Theorem 4]{Liu17_massive1}):
\begin{align}\label{eqn:asymptotic perfect detection}
\lim\limits_{M\rightarrow \infty} P_{t,n}^{{\rm MD}}(M)= \lim \limits_{M\rightarrow \infty} P_{t,n}^{{\rm FA}}(M)=0, ~~~ \forall t,n.
\end{align}This important result implies that in a massive MIMO system, in which $M$ can be larger than $100$, the AMP-based grant-free access scheme is able to detect device activity with extremely high accuracy in the massive IoT connectivity systems.

\begin{figure}[t]
  \centering
  \includegraphics[width=10cm]{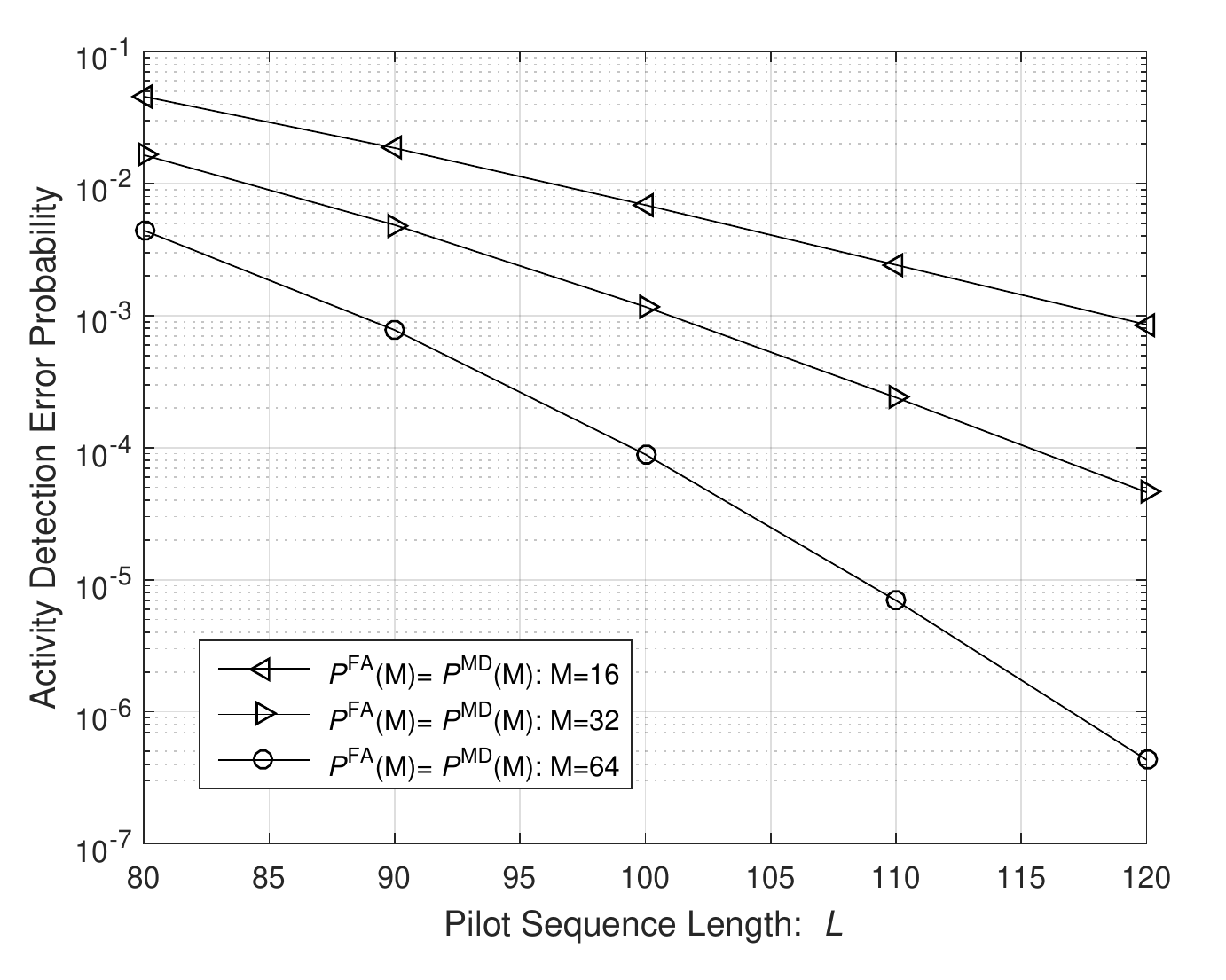}
  \caption{Probabilities of missed detection and false alarm versus pilot sequence length $L$ with massive MIMO.}\label{MD_FL_MMV}
\end{figure}

\begin{example}\label{example3}
Here we provide a numerical example to show the power of massive MIMO for massive IoT connectivity, under the same setup that is used in Examples \ref{example1} and \ref{example2}. Fig.~\ref{MD_FL_MMV} shows the probabilities of missed detection and false alarm (which are made equal by adjust the detection threshold) versus pilot sequence length $L$, with $M=16,32,64$ antennas at the BS. Here, similar to Example \ref{example2}, we define the system-level missed detection and false alarm probabilities as $P^{{\rm MD}}(M)=\sum_{n=1}^NP_n^{{\rm MD}}(M)$ and $P^{{\rm FA}}(M)=\sum_{n=1}^NP_n^{{\rm FA}}(M)$, where $P_n^{{\rm MD}}(M)$ and $P_n^{{\rm FA}}(M)$ denote the missed detection and false alarm probabilities of device $n$ achieved by AMP after its convergence. As compared to Fig.~\ref{MD_FL_SMV}, it is observed that even with $M=16$ antennas at the BS, both the missed detection and false alarm probabilities can be driven down to $10^{-3}$ when the pilot sequence is $L=120$, several orders of magnitude lower than the SMV case with the same $L$.
\end{example}

This article mainly focuses on the device activity detection performance under the grant-free access scheme. However, as shown in Fig. \ref{Grant Free}, besides device activity detection, channel estimation is performed as well via the metadata; moreover, data also should be decoded. Fortunately, the state evolution of the AMP algorithm enables us to characterize the channel estimation performance analytically, thus making it possible to quantify the user achievable rate with the effect of device activity detection into consideration \cite{Liu17_massive2}. Readers interested in information-theoretical studies on the capacity of the massive IoT connectivity systems with randomly active devices (also known as many-access channel) can refer to \cite{guo3,yu_ITA}. These references provide a justification for our proposed strategy to first detect the user activity via preambles then decode the user messages, i.e., the grant-free access scheme shown in Fig. \ref{Grant Free}.

\section{AMP-based Device Activity Detection with Embedded Information}
The AMP algorithm is introduced above for device activity detection. In this section, we
show how a modified version of AMP may be used for non-coherent detection of
information bits embedded in the pilot transmission. Although the two-phase grant-free access scheme shown in Fig. \ref{Grant Free} works very well for most of the cases when the user messages are of moderate and large size \cite{Liu17_massive2,guo3,yu_ITA}, as discussed at the end of last section, the strategy discussed in this section can be an effective alternative in the special case when very short messages (1 or several bits) are transmitted.

\subsection{Motivation}

In many applications the amount of data to be transmitted per block   may comprise only a small number of information bits, or even a single bit.  This situation is particularly common in control signaling, where the message  may contain acknowledgment (ACK/NACK) bits in a retransmission protocol, or simply a   concise request for a particular kind of response from the BS.

The transmission of extremely short packages is a challenging problem from two perspectives. First, fundamentally, the protection of very short packets
against transmission errors is very expensive. For a single bit,   repetition coding is the only possible strategy and for short blocks, block codes with low coding gains must be used. Second, as only error probability matters, capacity is an irrelevant metric. In fact, for extremely short blocks  even finite-block-length information theory becomes  inapplicable as the corresponding bounds and approximations  are too loose to be of  practical value.

As an aside, it is noteworthy that most academic work tend to deal with the transmission of long coded blocks, with   Shannon capacity as the primary performance metric. Contrarily, much of the effort invested in  standardization and system design is concerned with the   transmission of   short data blocks on the control plane, for which Shannon capacity  is, mostly, an illegitimate performance measure.  An explanation for this situation might be that digital transmission on the control plane is too hard to model and tackle with rigorous information theory: there   is no  ``Shannon theory'' available for its analysis -- whereas in contrast, established recipes are available for capacity   analysis of long-block transmission. A contributing reason might also be that  many academic researchers simply are unaware of the importance and the magnitude of the problem.

There are practical solutions for transmission of a single bit of control information. For example, \cite{piggy} considers the joint transmission of linearly coded payload data and a single ``additional bit''.  The transmitter uses the additional bit, through a one-to-one mapping,  to select one out of two possible codebooks  for the encoding of the payload. The receiver uses a  fast algorithm  to detect which codebook that was used, such that the additional bit can be detected before attempting to decode the payload data.

\subsection{Algorithm Design}

In the context of grant-free random access with non-orthogonal pilots, the main focus of our discussion, a small number, say $J$, of bits $b_1,\ldots,b_J$ may be encoded as follows \cite{Erik17,Erik18}. Each terminal is assigned, a priori, $2^J$ distinct, typically non-orthogonal, pilots. Upon transmission, the terminal uses the bits $\{b_i\}$ to select one of these $2^J$ pilots; specifically, it selects pilot number $1+b_1+2b_2+4b_3+\cdots+2^{J-1}b_J$ -- which depending on the bits $\{b_i\}$ ranges from 1 to $2^J$. The BS detects ``activity'' using the AMP algorithm -- but now, importantly, ``activity'' means the combination of the event that a particular terminal is active, and that a particular string of $J$ bits is being communicated. One may think of the
resulting communication scheme as non-coherent transmission.

The analytical model for device activity detection with embedded information in a massive MIMO system is given by:
\begin{align}\label{eqn:embedded information}
\mv{Y}=\bar{\mv{A}}\bar{\mv{X}}+\mv{Z},
\end{align}where $\bar{\mv{A}}=[\mv{a}_{1,1},\cdots,\mv{a}_{1,2^J},\cdots,\mv{a}_{N,1},\cdots,\mv{a}_{N,2^J}]\in \mathbb{C}^{L\times 2^JN}$ denotes the collection of all the $2^JN$ pilots that can be used by the devices, and $\bar{\mv{X}}=[\bar{\mv{x}}_{1,1},\cdots,\bar{\mv{x}}_{1,2^J},\cdots,\bar{\mv{x}}_{N,1},\cdots,\bar{\mv{x}}_{N,2^J}]^T\in \mathbb{C}^{2^JN\times M}$ denotes the collection of all the $2^JN$ effective channels of the devices. Specifically, the effective channel is modeled as $\bar{\mv{x}}_{n,i}=\alpha_{n,i}\mv{h}_n$, $n=1,\cdots,N$ and $i=1,\cdots,2^J$, where
\begin{align}\label{eqn:device activity with embedded information}
\alpha_{n,i}=\left\{\begin{array}{ll}1, & {\rm if ~ user} ~ n ~ {\rm is ~ active ~ and ~ its} ~ i{\rm th} ~ {\rm pilot ~ is ~ used}, \\ 0, & {\rm otherwise},\end{array} \right.
\end{align}As compared to (\ref{eqn:received signal multiple antenna}) for sole device activity detection, the dimensions of sensing matrix $\bar{\mv{A}}$ and effective channels $\bar{\mv{X}}$ are enlarged by a factor of $2^J$ to embed $J$ bits information.

The AMP for device activity detection in the form described above in principle could be directly applied to this problem as it stands. However, significantly, it is suboptimal because the BS knows \emph{a priori} that among the $2^J$ pilots assigned to each terminal, only one can be active at a time, i.e., if $\alpha_{n,i}=1$, then $\alpha_{n,j}=0$, $\forall j\neq i$.

Here we discuss the modified AMP algorithm for joint detection of user activity and embedded information bits, as proposed in \cite{Erik17}.  For conciseness of the exposition we focus here on the case of a single embedded bit $b$ (for which we omit
the index) -- i.e., $J=1$; then each user is assigned one out of two unique, but generally non-orthogonal pilot sequences. The modification of the AMP should introduce the constraint that out of the
two possible pilots, at most one may be transmitted at at a time; the
possible options are that either none of these pilots are sent (device silent), the
first one is sent (device active and communicates ``0''), or the second one is sent (device active and communicates ``1''). The overarching idea is to modify the AMP denoiser function, $\eta_{t,n}(\cdot)$,  to take into this constraint into account.

In more details, similar to (\ref{eq:stateqmm}), let $\hat{\mv{x}}_{n,1}=\bar{\mv{x}}_{n,1}+\mv{\Sigma}^{\frac{1}{2}}\mv{v}_n$ and $\hat{\mv{x}}_{n,2}=\bar{\mv{x}}_{n,2}+\mv{\Sigma}^{\frac{1}{2}}\mv{v}_n$ be the two vectors associated with the two possible pilots (for information bit ``0''  respectively ``1'') for device $n$; we omit the iteration index $t$ of the AMP algorithm here for brevity. The statistical characterization of $\hat{\mv{x}}_{n,1}$ and $\hat{\mv{x}}_{n,2}$ is
\begin{align}
  \hat{\mv{x}}_{n,i} & \sim  \left\{\begin{array}{ll} \mathcal{CN}(\mv{0}, \beta_n\mv{I} + \mv{\Sigma} ), & {\rm if} ~ \alpha_{n,i}=1, \\
 \mathcal{CN}(\mv{0}, \mv{\Sigma} ), & {\rm if} ~ \alpha_{n,i}=0, \end{array} \right. ~~~ i=1,2.
\end{align}
Based on these characterizations, we construct the following likelihood ratios:
\begin{align}
\lambda_{n,i} & = \frac{ p( \hat{\mv{x}}_{n,i}| \alpha_{n,i}=1) }{ p(\hat{\mv{x}}_{n,i}| \alpha_{n,i}=0) } =
\frac{ |\mv{\Sigma}  |}{|\beta_n\mv{I} +\mv{\Sigma}|} \exp\left(
-  \hat{\mv{x}}_{n,i}^H ( ( \beta_n\mv{I} +\mv{\Sigma} )^{-1}  - \mv{\Sigma}^{-1} )  \hat{\mv{x}}_{n,i}
\right).
\end{align}
We now re-work the denoiser, such that
in each time update, the constraint is taken into account that at most one of the vectors
$ \bar{\mv{x}}_{n,1}$ and $\bar{\mv{x}}_{n,2}$ can be nonzero.
Suppose that device $n$ is detected to be active. In principle, a comparison of $\lambda_{n,1}$ and $\lambda_{n,2}$ to a threshold would yield  a hypothesis test, that  could be used to discriminate between the two possibilities $\alpha_{n,1}=1$ and $\alpha_{n,2}=1$ or equivalently $b=0$ and $b=1$; one of $\bar{\mv{x}}_{n,1}$ and $\bar{\mv{x}}_{n,2}$ could then be set to zero based on the outcome of this test.
In this process, taking a soft decision as given in (\ref{eqn:soft thresholding denoiser MMV}) instead is preferable in order to avoid taking premature incorrect decisions on the embedded bits, which may propagate to subsequent iterations.
Experimentation in \cite{Erik17} has shown that a good heuristic is to use soft decision obtained by taking the original soft thresholding denoiser function given in (\ref{eqn:soft thresholding denoiser MMV}) and multiplying the denoisers for $\bar{\mv{x}}_{n,1}$ and $\bar{\mv{x}}_{n,2}$ by $\gamma( \lambda_{n,1}/(\lambda_{n,1}+\lambda_{n,2}))$ and $\gamma(\lambda_{n,2}/(\lambda_{n,1}+\lambda_{n,2}))$, respectively, where $\gamma(x)=1/(e^{-c(x-0.5)}+1)$ is a modified sigmoid function with its inflection point at $x=0.5$, 
where $c$ is a parameter to control the sharpness of the sigmoid function. The intuition is that the larger the likelihood ratio $\lambda_{n,1}$ is relative to $\lambda_{n,2}$, the more
likely it is that $\alpha_{n,1}=1$ or $b=0$ and the closer the weight for $\bar{\mv{x}}_{n,1}$
is to unity, and the closer the weight for $\bar{\mv{x}}_{n,2}$ is to zero.
This way, the effect of the denoiser on $\bar{\mv{x}}_{n,1}$ is similar to the effect of the soft thresholding (\ref{eqn:soft thresholding denoiser MMV}) as used in the original AMP algorithm solely for  device activity detection, whereas on $\bar{\mv{x}}_{n,2}$ is instead
pushed
down towards zero.  A similar interpretation holds for the opposite case when
$ \lambda_{n,1} < \lambda_{n,2}$.

Importantly, while the modified denoiser outlined here yields good results in numerical experiments, it is not optimal in any known sense. Research opportunities are available to find improved denoisers that can make a better utilization of the constraint that at most one of $\bar{\mv{x}}_{n,1}$ and $\bar{\mv{x}}_{n,2}$ can be nonzero. An extension of the modified AMP denoiser to the case of multiple embedded bits is available \cite{Erik18}.

A final remark is that the embedding of one or several information bit(s) of course incurs an expense of storage of more pilot sequences at the device and at the BS.  Also for given coherence block length, more resources need be dedicated to pilot transmission in order to maintain the same error probability performance. Yet, in the case to transmit very short messages, the embedding scheme has been shown to be efficient compared to conventional scheme consisting of pilot-based channel estimation (using the sparsity/AMP-based techniques proposed here) followed by coherent detection \cite{Erik18}.

\section{Other Compressed Sensing Techniques for Device Activity Detection}\label{sec:Other Compressed Sensing Techniques for Device Activity Detection}

Besides AMP, researchers with diverse backgrounds have developed
many other powerful algorithms to reconstruct sparse signals from low dimensional linear measurements as given in (\ref{eqn:received signal single antenna}) and (\ref{eqn:received signal multiple antenna}). These compressed sensing algorithms can also be leveraged in our considered massive IoT connectivity setting for device activity detection, e.g., coded slotted ALOHA.

\begin{figure}[t]
  \centering
  \includegraphics[width=10cm]{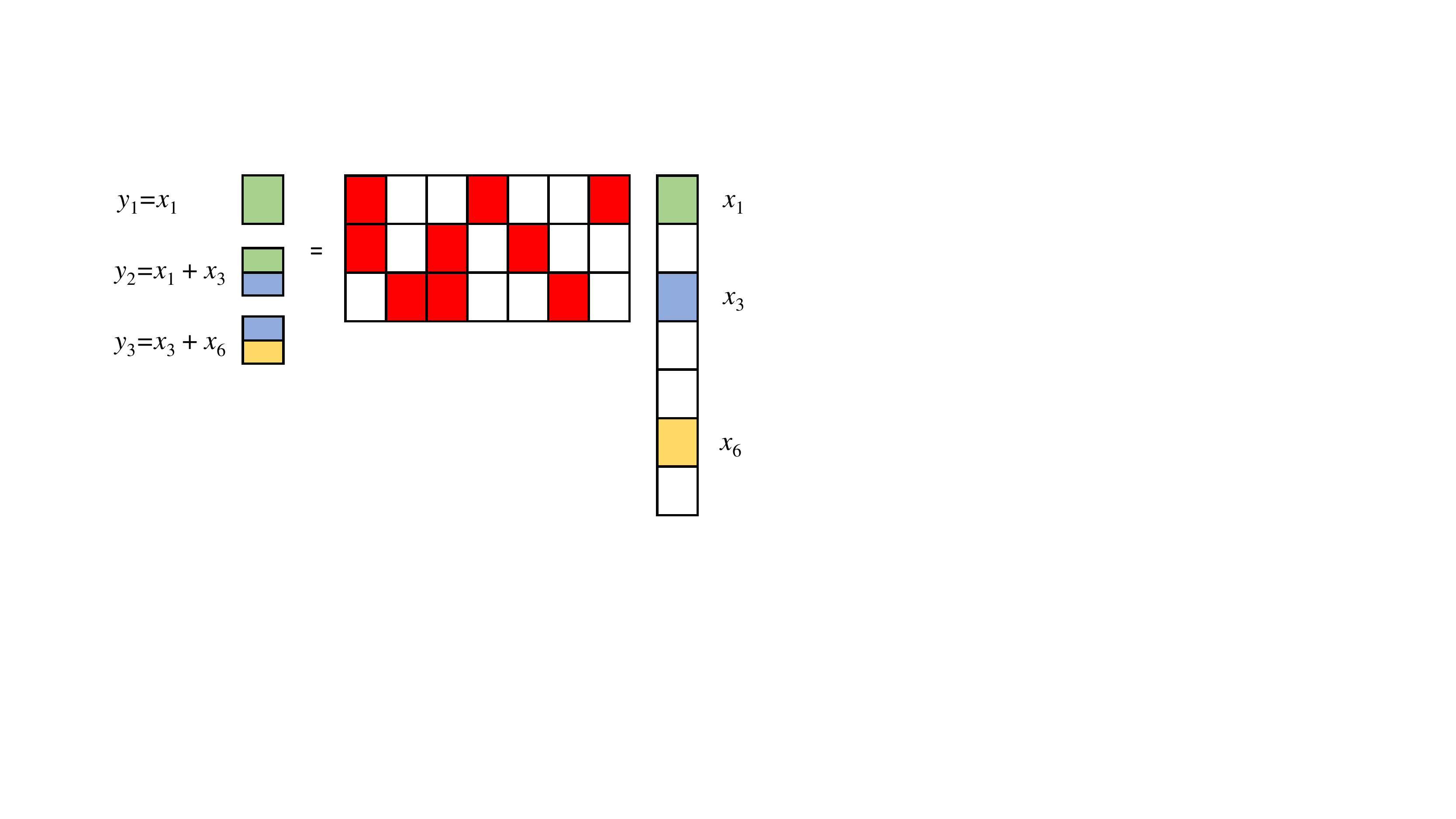}
  \caption{Example to illustrate the philosophy of the sparse-graph based compressed sensing algorithm \cite{Kannan15}. In this example, $\mv{x}$ is a 3-sparse signal, in which the zero entries are colored in white, while the non-zero entries, i.e., $x_1$, $x_3$, and $x_6$, are colored in green, blue, and yellow, respectively. Moreover, the sensing matrix $\mv{A}$ is sparse, in which the zero entries are colored in white, while the non-zero entries are colored in red. $x_1$ is detected from $y_1$. Then, $x_1$ is subtracted from $y_2$ so that $x_3$ is detected. Similarly, $x_6$ is detected after $x_3$ is removed from $y_3$.}\label{Sparse Graph}
\end{figure}

One powerful algorithm of low complexity is the so-called sparse-graph based compressed sensing algorithm \cite{Kannan15}, where the sensing matrix $\mv{A}$ is designed by sparsifying each row of the measurement matrix with zero patterns guided by sparse-graph codes. The reason for such a sparse sensing matrix design is to disperse the signal into single-tons that only contain one non-zero element in $\mv{x}$ and peel them off from multi-tons that contain two or more non-zero elements in $\mv{x}$ such that they can become single-tons.

Fig.~\ref{Sparse Graph} gives a simple example to briefly illustrate how this algorithm works to recover $\mv{x}$ in the ideal case without noise in (\ref{eqn:received signal single antenna}), in which the dimensions of $\mv{x}$ and $\mv{y}$ are $N=7$ and $L=3$, respectively, and $x_1$, $x_3$, $x_{6}$ are non-zero entries in $\mv{x}$. Moreover, the sensing matrix is
\begin{align}\label{eqn:sparse_graph}
\mv{A}=\left[\begin{array}{ccccccc} 1 & 0 & 0 & 1 & 0 & 0 & 1 \\ 1 & 0 & 1 & 0 & 1 & 0 & 0 \\ 0 & 1 & 1 & 0 & 0 & 1 & 0  \end{array} \right].
\end{align}Due to the sparsity in both $\mv{A}$ and $\mv{x}$, in the received signal $\mv{y}=[y_1,y_2,y_3]^T$, $y_1$, $y_2$, and $y_3$ only contain information about $x_1$, $x_1+x_3$, and $x_3+x_5$, respectively. Thus, $x_1$ can be detected from the single-ton $y_1$. Next, $x_1$ is removed from $y_2$ and $y_3$, which become single-tons so that $x_3$ and $x_6$ can be decoded.
Note that the effect of the channels is not taken into account in this example.

Density evolution, a powerful tool in modern coding theory, tracks the average density of remaining edges that are not decoded after a fixed number of peeling iteration. The convergence of the above algorithm is guaranteed by showing the convergence of the density evolution towards zero.

We remark that the above ``successive interference cancellation'' procedure is the principle of coded slotted ALOHA, a powerful multiuser access scheme in which the active devices transmit replicas of their packets in randomly chosen slots that contain both metadata (i.e., pilot sequences) and data. A successful detection of a packet replica in some slot enables removal of the related replicas from the slots in which they occur. This, in turn, lowers the number of colliding packets in the affected slots and boosts their detection probability, instigating new rounds of successive interference cancellation etc. If a single user packet can be detected in a slot, then the entries in $\mv{x}$ denote packets of active users, and entries in the sparse sensing matrix $\mv{A}$ denote the choice of the slots where the packets are repeated. On the other hand, the possibility to decode multiple user packets in a slot is also discussed in \cite{PSLP2014,WSPT2015} to improve the detection performance.

We remark that besides AMP and the sparse-graph based algorithm, many powerful compressed sensing algorithms exist in the literature, including LASSO \cite{Tibshirani96}, Orthogonal Matching Pursuit (OMP) \cite{Tropp07}, and so on. Further, the group-sparsity in the MMV model (\ref{eqn:received signal multiple antenna}) can also be utilized in LASSO \cite{Jacob09}, in which the $\ell_1/\ell_2$ penalty, i.e., the sum of $\ell_2$ norm penalty, is used to promote the desired sparsity pattern. The potential to apply these advanced compressed sensing techniques for user activity detection has been discussed in \cite{C-RAN,zhu,wunder_5G}.
It would be of great significance to investigate
which compressed sensing algorithm is best suited for device activity detection in the massive IoT connectivity setting, in terms of the complexity of pilot sequence design, the pilot sequence length required to achieve reasonable device detection accuracy, the corresponding missed detection and false alarm probabilities performance and channel estimation performance, etc.

\section{Conclusions}
A key feature of the future IoT network is the massive number of devices, e.g., sensors, actuators, and etc., each with sporadic data traffic. Facilitating the data transmission from so many IoT devices with extremely low latency poses plenty of new research challenges to the signal processing community. To embrace the upcoming era of IoT, this article advocates a grant-free access scheme that mitigates the delay arising from the contention resolution in the current random access scheme, and outlines a compressed sensing based approach for device activity detection to enable the grant-free access scheme to work. Most notably, the massive MIMO technology, originally proposed for improving the spectrum efficiency of human-type communications, is able to boost the device activity detection accuracy remarkably for massive IoT connectivity as well, with the aid of the MMV-based AMP algorithm. We have also discussed about the potential to decode some short messages along with the device activity detection process.

\begin{IEEEbiography}{Liang Liu} (lianguot.liu@utoronto.ca) received the B.Eng degree from the Tianjin University, China, in 2010, and the Ph.D degree from the National University of Singapore in 2014. He is currently a Research Fellow in the Department of Electrical and Computer Engineering at National University of Singapore. Before that, he was a Postdoctoral Fellow at the University of Toronto, Toronto, Ontario, Canada. His research interests include energy harvesting, convex optimization, and machine-type communications in 5G. He was the recipient of the IEEE Signal Processing Society Young Author Best Paper Award, 2017, and a Best Paper Award from IEEE WCSP 2011.
\end{IEEEbiography}

\begin{IEEEbiography}{Erik G. Larsson} (erik.g.larsson@liu.se) is Professor at Link\"oping University, Sweden.
He   co-authored   \emph{Fundamentals of Massive MIMO}
 (Cambridge, 2016) and \emph{Space-Time
Block Coding for Wireless Communications} (Cambridge, 2003).
Recent service includes membership of  the IEEE Signal Processing Society
Awards Board (2017--2019),  and the  \emph{IEEE Signal Processing Magazine}
editorial board  (2018--2020).
He received the  \emph{IEEE Signal Processing Magazine} Best
Column Award twice, in 2012 and 2014, the IEEE ComSoc Stephen
O. Rice Prize in Communications Theory  2015,
 the IEEE ComSoc Leonard G. Abraham
Prize  2017 and the IEEE ComSoc Best Tutorial Paper Award 2018.
He is a   Fellow of the IEEE.
\end{IEEEbiography}

\begin{IEEEbiography}{Wei Yu} (weiyu@comm.utoronto.ca) received Ph.D. degree in electrical engineering from Stanford University in 2002, and is Professor and Canada Research Chair in Information Theory and Wireless Communications at the University of Toronto, Canada. He received the IEEE Signal Processing Society Best Paper Award in 2008 and 2017. He currently serves on the Board of Governors of the IEEE Information Theory Society, and serves as Chair of the Signal Processing of Communications and Networking Technical Committee of the IEEE Signal Processing Society. He is a Fellow of IEEE and a Fellow of Canadian Academy of Engineering.
\end{IEEEbiography}

\begin{IEEEbiography}{Petar Popovski} (petarp@es.aau.dk) is a professor at Aalborg University, Denmark. He received his Dipl.-Ing./ Mag.-Ing. in communication engineering from Sts. Cyril and Methodius University in Skopje, R. of Macedonia, and his Ph.D. from Aalborg University. He received an ERC Consolidator Grant (2015), the Danish Elite Researcher award (2016), IEEE Fred W. Ellersick prize (2016) and IEEE Stephen O. Rice prize (2018). He is currently an Area Editor for IEEE Transactions on Wireless Communications and a Steering Board member of IEEE SmartGridComm. His research interests are in wireless communications/networks and communication theory.
\end{IEEEbiography}

\begin{IEEEbiography}{\v{C}edomir Stefanovi\'{c}} (cs@es.aau.dk) received the Dipl.-Ing., Mr.-Ing., and Dr.-Ing. degrees in electrical engineering from the University of Novi Sad, Serbia. He is currently an associate professor at the Department of Electronic Systems, Aalborg University, Denmark. He was and currently is involved in a number of national and EU projects on IoT and 5G communications. Hi is an editor for IEEE Internet of Things Journal. His research interests include communication theory, wireless and smartgrid communications.
\end{IEEEbiography}

\begin{IEEEbiography}{Elisabeth de Carvalho} (edc@es.aau.dk) received a Ph.D. in electrical engineering from
Telecom ParisTech, France. After her Ph.D. she was a post-doctoral fellow
at Stanford University, USA and then worked in industry in the field of DSL
and wireless LAN. Since 2005, she has been an associate professor at
Aalborg University where she has led several research projects in wireless
communications. Her main expertise is in signal processing for MIMO
communications with recent focus on massive MIMO including channel
measurements, channel modeling, beamforming and protocol aspects. She is
a coauthor of the text book ``A practical guide to the MIMO radio channel''.
\end{IEEEbiography}

\end{document}